\documentstyle[prb,aps,epsfig]{revtex}
\begin{document}

\title{In-plane upper critical field anisotropy in Sr$_2$RuO$_4$ and CeIrIn$_5$}
\author{D.F. Agterberg}
\address{Department of Physics, University of Wisconsin-Milwaukee, Milwaukee, WI 53201}

\maketitle
\begin{abstract}
Experiments on tetragonal Sr$_2$RuO$_4$ and CeIrIn$_5$ indicate the presence of superconductivity with a
multi-component superconducting order parameter. Such an order parameter should exhibit an in-plane anisotropy in
the upper critical field near the superconducting transition temperature that does not occur for single-component
superconductors. Here, this anisotropy is determined from microscopic calculations for arbitrary gap functions. It
is shown that this anisotropy is generally not small and, in some cases, independent of impurity scattering.
Furthermore, this anisotropy is calculated for many detailed microscopic models of Sr$_2$RuO$_4$. For these models
this anisotropy is found to be large, which is in sharp contrast to the small anisotropy observed experimentally.
However, an accidental cancellation of the anisotropy for gaps on different Fermi surface sheets can lead to a
result that is consistent with experiment.
\end{abstract}

\vglue 0.5 cm

One of the central issues in the field of unconventional superconductivity is the identification of the order
parameter symmetry. This issue can usually only be addressed by determinations of the phase of the order parameter
(such as through Josephson junction experiments \cite{van95}). However, in the case that the Cooper pair wave
functions have more than one complex degree of freedom, other experimental signatures can reveal the pairing
symmetry. In this paper one of these signatures, the upper critical field anisotropy for fields in the basal
plane, is explored for tetragonal superconductors. This anisotropy was established by Gor'kov on phenomenological
grounds \cite{gor87}. Such an anisotropy, for which $dH_{c2}/dT|_{T=T_c}$ is not equal for the magnetic field
applied along the $(1,0,0)$ (along the crystallographic ${\bf a}$ axis) and the $(1,1,0)$ directions, cannot occur
for superconducing order parameters that have only one complex degree of freedom (for example conventional
$s$-wave superconductors or $d_{x^2-y^2}$-wave superconductors). Consequently, this provides a clear test of the
pairing symmetry. For tetragonal materials, there exist two possible pairing states that exhibit this property.
One is the spin triplet $E_u$ representation which can be described a gap function of the form (note that this is
not the most general form) ${\bf d}({\bf k})=\hat{z}[\eta_1 f_x({\bf k})+\eta_2 f_y({\bf k})]$, where in the
simplest case (for a cylindrical Fermi surface) $f_x({\bf k})=k_x$ and $f_y({\bf k})=k_y$. Such a gap function is
believed to describe the pairing state in Sr$_2$RuO$_4$ (see below)\cite{mae01,ric98}. The other pairing state is
the spin-singlet $E_g$ representation which can be described by a gap function of the form $\Psi({\bf k})=
\eta_1f_{xz}({\bf k})+\eta_2 f_{yz}({\bf k})$, where $f_{xz}({\bf k})=k_xk_z$ and $f_{yz}({\bf k})=k_yk_z$ in the
simplest case. In this article, the anisotropy is determined within microscopic theories with arbitrary gap
functions. It is shown that the anisotropy is generally not small and consequently easily observable. Furthermore,
it is shown that for the $E_g$ representation and for some theories for the $E_u$ representation, this anisotropy
is impurity independent. Application of these results to Sr$_2$RuO$_4$ and CeIrIn$_5$ are discussed and detailed
calculations for Sr$_2$RuO$_4$ are also presented. Prior to discussing the anisotropy, an overview of the
superconductivity in Sr$_2$RuO$_4$ and CeIrIn$_5$ is given.

The oxide Sr$_2$RuO$_4$ has a structure similar to high $T_c$ materials and was observed to be superconducting by
Maeno {\it et al.} in 1994 \cite{mae94}. It has been established that this superconductor is not a conventional
$s$-wave superconductor: NQR measurements show no indication of a Hebel-Slichter peak \cite{ish97} in $1/T_1T$,
and $T_c$ is strongly suppressed below the maximum value of 1.5 K by non-magnetic impurities \cite{mac98}. More
recent experiments indicate an $E_u$ odd parity gap function of the form ${\bf d}({\bf k})=\hat{z}[\eta_1 f_x({\bf
k}) +\eta_2 f_y({\bf k})]$. The Knight shift measurements of Ishida {\it et al.} \cite{ish98} reveal that the spin
susceptibility is unchanged upon entering the superconducting state; this is consistent with $p$-wave
superconductivity (as predicted by Rice and Sigrist \cite{ric95} and Baskaran \cite{bas96}). Furthermore, these
measurements were conducted with the applied field in the basal plane. Since the orientation of the gap function
(that is ${\bf d}$) is orthogonal to the spin projection of the Cooper pair \cite{sig91}, these measurements are
consistent with the gap function aligned along the $\hat{z}$ direction. The $\mu$SR experiments of Luke {\it et.
al.} have revealed spontaneous fields in the Meissner state \cite{luk98}. This indicates that the superconducting
order parameter must have more than one component \cite{sig91,luk98}. This leads naturally to the conclusion that
the superconducting gap function has the form ${\bf d}({\bf k})=\hat{z}(f_x\pm i f_y)$. This conclusion is
consistent with measurements of the field distribution of the vortex lattice \cite{kea00}, Josephson tunnelling
experiments \cite{jin00}, and point-contact measurements \cite{lau00}. One immediate consequence of such a gap
function is that the resulting low energy excitation spectrum is gapless in the clean limit. However, recent
experiments indicate this is not the case. Specific heat measurements \cite{nis99}, NQR measurements \cite{ish00},
and cavity penetration depth measurements \cite{bon00} all indicate the presence of nodes in the superconducting
gap function. This has prompted a recent series of proposals that the gap function has the form ${\bf d}({\bf
k})=\hat{z}[\eta_1 f_x({\bf k})+\eta_2 f_y({\bf k})]$ where $f_x({\bf k}_0)=f_y({\bf k}_0)=0$ for a set of points
${\bf k}_0$ that form a line on the Fermi surface \cite{has00,gra00,mak00} or are very anisotropic
\cite{agt97,miy99,kur01,zhi01}. These gap functions are all formally of the same $E_u$ symmetry. The fact that
these theories all have the same symmetry implies that there exist experimental signatures that will be common to
all of them. In particular, the upper critical field anisotropy discussed above should exist. Recently, such an
anisotropy has been observed by Mao {\it et al.}\cite{mao00}. It was determined that $1-
H_{c2}^{(100)}/H_{c2}^{(110)}=0.03$ at a temperature of $0.35 K$ ($\approx T_c/5$), $1-
H_{c2}^{(100)}/H_{c2}^{(110)}$ was observed to change sign as $T\rightarrow T_c$, and appears to remain small for
$T$ near $T_c$. In this paper, the anisotropy near $T_c$ will be calculated for a variety of models.

Very recently, it has been reported that $\mu$SR experiments of Heffner  {\it et. al.} have revealed spontaneous
fields in the Meissner state of tetragonal CeIrIn$_5$ \cite{pet01,hef01}. Just as in the case for Sr$_2$RuO$_4$
above, this indicates that the superconducting order parameter must have more than one component
\cite{sig91,luk98}. This leads naturally to the conclusion that the superconducting gap function is either of
$E_u$ symmetry of of $E_g$ symmetry. Consequently, both pairing symmetries are considered in this article.

The free energy for the $E_u$ and $E_g$ representations of $D_{4h}$ is given by \cite{sig91}
\begin{eqnarray}
f=&-\alpha |\vec{\eta}|^2+\beta_1|\vec{\eta}|^4/2+
\beta_2(\eta_1\eta_2^*-\eta_2\eta_1^*)^2/2 \nonumber \\
& +\beta_3|\eta_1|^2|\eta_2|^2 +
\kappa_1|D_x\eta_1|^2+|D_y\eta_2|^2 \nonumber \\
& +\kappa_2(|D_y\eta_1|^2+ |D_x\eta_2|^2) \label{eq1} \\ &+ \kappa_5(|D_z\eta_1|^2+ |D_z\eta_2|^2) \nonumber
\\ & +\kappa_3[(D_x\eta_1)(D_y\eta_2)^*+h.c.] \nonumber
\\ & +
\kappa_4[(D_y\eta_1)(D_x\eta_2)^*+h.c.] +h^2/(8\pi). \nonumber
\end{eqnarray}
where $D_j=\nabla_j-\frac{2ie}{\hbar c} A_j$, ${\bf h}=\nabla\times {\bf A}$, and ${\bf A}$ is the vector
potential. The stable homogeneous solutions are easily determined \cite{sig91}. There are three phases: (a)
$\vec{\eta}=(1,i)/\sqrt{2}$ ($\beta_2>0$ and $\beta_2>\beta_3/2$), (b) $\vec{\eta}=(1,0)$ ($\beta_3>0$ and
$\beta_2<\beta_3/2$), and (c) $\vec{\eta}=(1,1)/\sqrt{2}$ ($\beta_3<0$ and $\beta_2<0$). The weak-coupling
approximation predicts that phase (a) is stable since this phase minimizes the number of nodes in the order
parameter [note that in exceptional circumstances phase (a) may be degenerate with phase (b) and (c) in the
weak-coupling limit, for example in the theory of Ref.~\cite{agt98-2} for $|\nu|=1$].

For an external magnetic field applied in the basal plane, the upper critical field can be determined and is found
to be \cite{gor87,sig91}
\begin{equation}
H_{c2}(\theta)=\frac{\alpha
hc}{e\sqrt{2\kappa_5}\sqrt{\kappa_2+\kappa_1-\sqrt{(\kappa_1-\kappa_2)^2\cos^2(2\theta)
+(\kappa_3+\kappa_4)^2\sin^2(2\theta)}}} \label{eq2}
\end{equation} where $\theta$ is the angle between the magnetic field and the crystallographic ${\bf a}$ axis.
From the above expression, the ratio of $H_{c2}$ for the field along the $(1,0,0)$ direction to that for the field
along the $(1,1,0)$ direction is given by
$H_{c2}^{(1,1,0)}/H_{c2}^{(1,0,0)}=\sqrt{(\kappa_1+\kappa_2-|\kappa_3+\kappa_4|)/2min(\kappa_2,\kappa_1)}$ where
$min(\kappa_2,\kappa_1)$ means the minimum of $\kappa_1$ and $\kappa_2$. Note that Sigrist has also determined an
additional anisotropy that occurs due to spin-orbit coupling \cite{sig00}. This additional anisotropy introduces a
correction to Eq.~\ref{eq2} that vanishes as $T\rightarrow T_c$, so it is not considered here.

Here, this anisotropy is determined in the weak-coupling approximation. This approximation is reasonable for
Sr$_2$RuO$_4$ since $T_c/T_F\approx 10^{-4}$, however it is not clear whether or not this approximation is valid
for CeIrIn$_5$. Using a gap function of the form ${\bf d}({\bf k})=\hat{z}[\eta_1 f_x({\bf k})+\eta_2 f_y({\bf
k})]$ for the $E_u$ representation and introducing isotropic impurity scattering within a $T$-matrix approximation
yields (the details will be given in a later publication)
\begin{eqnarray}
\kappa_1=&\frac{\pi N(0)}{2}\sum_{n\ge0}\frac{1}{(\omega_n+\Gamma)^3}[\langle f_x^2 v_x^2\rangle+
\frac{\Gamma}{\omega_n}\langle f_x v_x\rangle^2] \nonumber\\
\kappa_2=&\frac{\pi N(0)}{2}\sum_{n\ge0} \frac{1}{(\omega_n+\Gamma)^3}\langle f_x^2 v_y^2\rangle \nonumber\\
\kappa_3=&\frac{\pi N(0)}{2}\sum_{n\ge0} \frac{1}{(\omega_n+\Gamma)^3}\times\nonumber\\& [\langle f_xf_y
v_xv_y\rangle+\frac{\Gamma}{\omega_n}
\langle f_xv_x\rangle^2]\nonumber \\
\kappa_4=&\frac{\pi N(0)}{2}\sum_{n\ge0} \frac{1}{(\omega_n+\Gamma)^3}\langle f_xf_y
v_x v_y\rangle\nonumber \\
\kappa_5=&\frac{\pi N(0)}{2}\sum_{n\ge0} \frac{1}{(\omega_n+\Gamma)^3}\langle f_x^2 v_z^2\rangle \nonumber
\end{eqnarray}
where $\omega_n=(2n+1)\pi T$, $N(0)$ is the density of states at the Fermi surface, $\Gamma=1/\tau_N$ is half of
the scattering rate, $v_i$ are the components of the Fermi velocity, and the brackets $\langle \rangle$ denote an
average over the Fermi surface. It has also been assumed that $\langle f_x^2+f_y^2 \rangle =1$. The terms
proportional to $\Gamma$ in $\kappa_1$ and $\kappa_3$ are vertex corrections (which have the same form as those
that arise in the calculation of current-current correlation functions). In the clean limit, this gives
\begin{equation}
\frac{H_{c2}^{(1,1,0)}}{H_{c2}^{(1,0,0)}}= \sqrt{\frac{\langle f_x^2v_x^2\rangle+\langle f_y^2v_x^2\rangle-2
|\langle f_xf_yv_xv_y\rangle|}{2 min(\langle f_x^2v_y^2\rangle,\langle f_x^2v_x^2\rangle)}}. \end{equation} For
$\kappa_2<\kappa_1$ and $\kappa_3+\kappa_4>0$, this ratio is independent of the impurity concentration. The
requirement that $\kappa_2<\kappa_1$ implies for the field along $(1,0,0)$ that $(\eta_1,\eta_2)=\eta({\bf
r})(1,0)$ [for $\kappa_2>\kappa_1$ $(\eta_1,\eta_2)=\eta({\bf r})(0,1)$] while the requirement that
$\kappa_3+\kappa_4>0$ implies that for the field along the $(1,1,0)$ direction that $(\eta_1,\eta_2)=\eta({\bf
r})(1,1)$ [if $\kappa_3+\kappa_4<0$ then $(\eta_1,\eta_2)=\eta({\bf r})(1,-1)$]. Note that as the magnetic field
is decreased, there will exist a second order phase transition due to a change in the order parameter structure
\cite{agt98}. For example, for the field along $(1,0,0)$ and for $\kappa_2<\kappa_1$ there will be a second order
transition with decreasing field where the component $(\eta_1,\eta_2)\propto (0,1)\ne 0$. Such a transition
reduces the number of nodes in the gap structure which is why it occurs.

For the $E_g$ representation using $\Psi({\bf k})=(\eta_1 f_{xz}+\eta_2 f_{yz})$ gives
\begin{eqnarray}
\kappa_1=&\frac{\pi
N(0)}{2}\sum_{n\ge0}\frac{1}{(\omega_n+\Gamma)^3}\langle f_{xz}^2 v_x^2\rangle \nonumber\\
\kappa_2=&\frac{\pi N(0)}{2}\sum_{n\ge0} \frac{1}{(\omega_n+\Gamma)^3}\langle f_{xz}^2 v_y^2\rangle \nonumber\\
\kappa_3=&\frac{\pi N(0)}{2}\sum_{n\ge0} \frac{1}{(\omega_n+\Gamma)^3}\times\nonumber\\& [\langle f_{xz}f_{yz}
v_xv_y\rangle\\
\kappa_4=&\frac{\pi N(0)}{2}\sum_{n\ge0} \frac{1}{(\omega_n+\Gamma)^3}\langle f_{xz}f_{yz}
v_x v_y\rangle\nonumber \\
\kappa_5=&\frac{\pi N(0)}{2}\sum_{n\ge0} \frac{1}{(\omega_n+\Gamma)^3}\langle f_{xz}^2 v_z^2\rangle \nonumber
\end{eqnarray}
The main difference with the $E_u$ representation is that no vertex corrections appear. For an arbitrary impurity
concentration this gives
\begin{equation}
\frac{H_{c2}^{(1,1,0)}}{H_{c2}^{(1,0,0)}}= \sqrt{\frac{\langle f_{xz}^2v_x^2\rangle+\langle f_{yz}^2v_x^2\rangle-2
|\langle f_{xz}f_{yz}v_xv_y\rangle|}{2 min(\langle f_{xz}^2v_y^2\rangle,\langle f_{xz}^2v_x^2\rangle)}}.
\end{equation}
This ration is independent of the impurity concentration. These results (and the more detailed calculations below)
indicate that the anisotropy is not generally small and should therefore be easily detected. It would be of
interest to look for this anisotropy in CeIrIn$_5$.

It is informative to determine the anisotropy for existing microscopic theories in the clean limit. Since no such
theories for CeIrIn$_5$ exist to date, the remainder of the analysis concentrates on Sr$_2$RuO$_4$. To do this the
electronic dispersion must be given as must the form of the gap function. Initially, consider theories that for
simplicity  use a cylindrical Fermi surface \cite{has00,gra00,mak00,miy99}. For a gap function with nodes in the
basal plane $(f_x,f_y)=[(k_x^2-k_y^2)k_x,-(k_x^2-k_y^2)k_y]$ \cite{has00,gra00},
$H_{c2}^{(1,1,0)}/H_{c2}^{(1,0,0)}=0.58$. For an anisotropic but fully gapped gap function:
$(f_x,f_y)=[\sin(k_x),\sin(k_y)]$ with $k_F=0.9\pi$ \cite{miy99}, $H_{c2}^{(1,1,0)}/H_{c2}^{(1,0,0)}=0.53$. These
estimates indicate that the anisotropy should be easily observable if the gap contains nodes in the basal plane or
is very anisotropic (these possibilities might be expected due to the recent experiments that indicate nodes in
the gap \cite{nis99,ish00,bon00}). This in contrast to the experimental results of Mao {\it et al.} \cite{mao00}.
A more careful investigation below indicates that the anisotropy can accidentally be hidden, even for theories
with nodes in the plane.

To consider more realistic models, a tight binding approach is used to describe the electronic dispersion near the
Fermi surface. Local-density approximation band-structure calculations reveal that the density of states near the
Fermi surface is due mainly to the four Ru $4d$ electrons in the $t_{2g}$ orbitals \cite{ogu95,sin95}. There is a
strong hybridization of these orbitals with the O $2p$ orbitals giving rise to antibonding $\pi^*$ bands. The
resulting bands have three quasi-2D Fermi surface sheets labelled $\alpha$, $\beta$ and $\gamma$ (see Fig. 1). To
model these Fermi surface sheets, the following tight-binding dispersions are used:
\begin{eqnarray}
\epsilon_{\gamma}=&\epsilon_{\gamma}^0-2t_{\gamma}(\cos k_x+\cos k_y)-4\tilde{t}_{\gamma}\cos k_x \cos k_y \nonumber\\
\epsilon_{\alpha,\beta}=&\epsilon_{\alpha,\beta}^0-t_{\alpha,\beta}(\cos k_x+\cos k_y)\pm
\sqrt{t_{\alpha,\beta}^2(\cos k_x -\cos k_y)^2+16 \tilde{t}^2_{\alpha,\beta}\sin^2 k_x\sin^2k_y}
\end{eqnarray}
In following tight binding values are used: $(\epsilon_{\gamma}^0,t_{\gamma},\tilde{t}_{\gamma})=(-0.4,0.4,0.12)$
for the $\gamma$ sheet and the values
$(\epsilon_{\alpha,\beta}^0,t_{\alpha,\beta},\tilde{t}_{\alpha,\beta})=(-0.3,0.25,0.0375)$ for the
$\{\alpha,\beta\}$ sheets \cite{maz97}. Table 1 shows the anisotropy ratio for gap functions found in various
theories for Sr$_2$RuO$_4$.

Note that for the $\alpha$ and $\beta$ bands the gap functions with $v_x$ replaced by $\sin k_x$ were not included
(as they were for the $\gamma$ band) . This is due to quasi one-dimensional nature of the $\alpha$ and $\beta$
sheets. In particular, $v_{x,\alpha}\propto \sin k_x[1+sgn(\cos k_x-\cos k_y)]/2\ne \sin k_x$ in the
one-dimensional limit. It is $v_x$, not $\sin k_x$, which would be the gap function for a one-dimensional nearest
neighbor pairing interaction. This is supported by the calculations of Kuroki {\it et al.} \cite{kur01} where the
$(f_x,f_y)=(v_x,v_y)$ basis functions on the $\alpha$ sheet of Ref. \cite{kur01} gives a good approximation to the
crib-shaped gap function found in this reference [the $(f_x,f_y)=(\sin k_x,\sin k_y)$ basis functions gives a
qualitatively different gap function, with gap maxima along the $(1,1,0)$ momentum direction]. A point of interest
is that $H_{c2}^{100}/H_{c2}^{110}<1$ for the all the $\gamma$ sheet gap functions and
$H_{c2}^{100}/H_{c2}^{110}>1$ for all the $\alpha$ and $\beta$ sheet gap functions. Consequently, the upper
critical anisotropy can cancel between these sheets of the Fermi surface. For example, using the gap function
$f_x=v_x$ and taking the magnitude of this gap function on the $\alpha$ and $\beta$ surfaces to be the same
($c_{\alpha}=c_{\beta}$) the gap magnitude ratio $c_{\gamma}/c_{\alpha}=6.7$ will lead to zero upper critical
field anisotropy (where the experimentally measured ratios
$N_{\gamma}(0):N_{\beta}(0):N_{\alpha}(0)=1.334:0.914:0.216$ and
$v_{F,\gamma}^2:v_{F,\beta}^2:v_{F,\alpha}^2=2.6:6.8:7.8$ were used \cite{ris98}). The same can be done for gap
functions with line nodes: for $f_x=v_y\sin k_x\sin k_y$, $c_{\gamma}/c_{\alpha}=1.7$ and for $f_x=(\cos k_x -\cos
k_y)v_x$, $c_{\gamma}/c_{\alpha}=55$. These estimates imply that for such an accidental cancellation to occur the
maximum gap on the $\gamma$ Fermi surface sheet must be larger than that on the $\{\alpha,\beta\}$ sheets. Note
that in principle all the gap functions listed in Table 1 can appear simultaneously in the pairing state of
Sr$_2$RuO$_4$ since they are all of the same symmetry.

The in-plane anisotropy in the upper critical field near $T_c$ has been calculated for microscopic models of the
$E_g$ and $E_u$ pairing symmetries of the tetragonal point group. It has been shown that this anisotropy should be
easily observable and may be present in CeIrIn$_5$. Also it is shown that for $E_g$ gap functions and for some
$E_u$ gap functions, this anisotropy is independent of impurity scattering. For Sr$_2$RuO$_4$, it is shown that an
accidental cancellation of this anisotropy can occur if the $\gamma$ sheet of the Fermi surface is largely
responsible for the superconductivity.

 I wish to thank V. Barzykin, L.P. Gor'kov, R.B. Joynt, Y. Maeno, M. Sigrist, and L. Taillefer
for informative discussions.

\begin{figure}
\epsfxsize=2.5 in \epsfbox{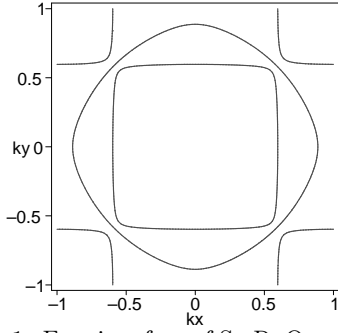} \caption{Fermi surface of Sr$_2$RuO$_4$ using the tight binding
parameters in this paper ($k_x$ and $k_y$ are in units $\pi/a$ where $a$ is the lattice spacing). Moving out from
the ${\bf k}=(0,0)$ point, first the $\beta$ sheet, then the $\gamma$ sheet, and finally the $\alpha$ sheet are
crossed.}
 \label{fig1}
\end{figure}

\begin{table}
\begin{tabular}{ccccc}
$f_x({\bf k})$ & $f_y({\bf k})$ & Fermi Surface&$\frac{H_{c2}^{100}}{H_{c2}^{110}}$& Impurity Dependent\\
\hline
$v_x$ & $v_y$ & $\gamma$ & 0.50&no\\
$(\cos k_x -\cos k_y)v_x$&$-(\cos k_x -\cos k_y)v_y$ & $\gamma$ & 0.86 &yes\\
$\sin k_x \sin k_y v_y$& $\sin k_x \sin k_y v_x$ & $\gamma$ & 0.31& yes \\
 $\sin k_x$ & $\sin k_y$ & $\gamma$ & 0.36& yes \\
  $(\cos k_x -\cos k_y)\sin k_x$& $(\cos k_x -\cos k_y)\sin k_y$& $\gamma$ & 0.51& yes \\
   $\sin^2 k_y \sin k_x$ & $\sin^2 k_x \sin k_y$ & $\gamma$ & 0.24 & no\\
    $v_x$ & $v_y$ & $\alpha$ & 2.0 &no\\
  $(\cos k_x -\cos k_y)v_x$&$-(\cos k_x -\cos k_y)v_y$ & $\alpha$ & 3.9 &yes\\
  $\sin k_x \sin k_y v_y$& $\sin k_x \sin k_y v_x$ & $\alpha$ & 1.2& yes \\
 $v_x$ & $v_y$ & $\beta$ & 2.5&no\\
  $(\cos k_x -\cos k_y)v_x$&$-(\cos k_x -\cos k_y)v_y$ & $\beta$ & 5.23 &yes\\
  $\sin k_x \sin k_y v_y$& $\sin k_x \sin k_y v_x$ & $\beta$ & 1.52& yes \\
\end{tabular}
\caption{Upper critical field anisotropy for various gap functions for Sr$_2$RuO$_4$. Note that the normalization
of the gap functions has not been included here. The column called Impurity Dependent states whether or not
$H_{c2}^{100}/H_{c2}^{110}$ depends upon impurity scattering.}\label{tab1}
\end{table}

\end{document}